\input{aipcheck}

\documentclass[
    ,final            
  ]
  {aipproc}

\layoutstyle{6x9}
\bibliography{biblio}

\begin{document}

\title{Modules identification by a Dynamical Clustering algorithm based on
chaotic R\"ossler oscillators}

\classification{89.75.-k, 05.45.Xt, 87.18.Sn}

\keywords      {Networks, Dynamical Clustering, Synchronization}

\author{Alessandro Pluchino}{
  address={Dipartimento di Fisica e Astronomia and INFN, Via S. Sofia 64, 95127 Catania, Italy}
}
\author{Vito Latora}
{
  address={Dipartimento di Fisica e Astronomia and INFN, Via S. Sofia 64, 95127 Catania, Italy}
}
\author{Andrea Rapisarda}
{
  address={Dipartimento di Fisica e Astronomia and INFN, Via S. Sofia 64, 95127 Catania, Italy}
}

\author{Stefano Boccaletti}
{
  address={CNR- Istituto dei Sistemi Complessi, Via Madonna del Piano, 10, 50019 Sesto Fiorentino (FI), Italy
     and the Italian Embassy in Tel Aviv, Trade Tower, 25 Hamered Street, Tel Aviv, Israel}
}

\begin{abstract}
A new dynamical clustering algorithm for the identification of modules in complex networks 
has been recently introduced \cite{BILPR}. 
In this paper we present a modified version of this algorithm based on a system of chaotic 
R\"ossler oscillators and we test its sensitivity 
on real and computer generated networks
with a well known modular structure. 
\end{abstract}

\maketitle


\section{Introduction}

An important property common to many 
networks is the presence of {\it modules} or 
{\it community structures}, that can be roughly defined
as subsets of network nodes {\it within} which the network connections are dense,
but {\it between} which they are sparser.
Since nodes belonging to tightly connected clusters of nodes are most likely
to have other properties in common, the detection of these structures
in complex networks is potentially very useful. 
\\
We have recently proposed a {\it dynamical clustering} (DC) method for the
modules identification based on the properties of a dynamical system 
associated to the graph \cite{BILPR}. 
Such a method combines topological and dynamical information
in order to devise an algorithm that is able to identify the
modular structure of a graph with a precision and a computational cost
($O(N^2)$) competitive with the best techniques based solely on the topology. 
The method is based upon the well-known phenomenon of synchronization
clusters of non identical phase oscillators \cite{boc02}, each one
associated to a node, and interacting through the edges of the
graph. Clusters of synchronized oscillators represent an
intermediate regime between global phase locking and full absence
of synchronization, thus implying a division of the whole graphs
into groups of elements which oscillate at the same (average)
frequency. The key idea is that, starting from a fully
synchronized state of the network, a dynamical change in the
weights of the interactions, retaining information of the
original betweenness distribution, yields a progressive
hierarchical clustering that fully detects modular structures.
\\
In this paper we implement our algorithm on a systems of
$N$ identical (three-dimensional) chaotic R\"ossler oscillators and 
we test the precision (sensitivity) obtained by the algorithm 
on real and computer generated networks with well known modular structures. 

\section{Dynamics of a weighted network of R\"ossler oscillators}

The dynamics of a network of $N$ coupled identical oscillators is described by:
\begin{eqnarray}
\dot{\bf x}_i&=&{\bf F}({\bf x}_i)-\sigma\sum_{j=1}^{N}G_{ij} {\bf H}[{\bf x}_i - {\bf x}_j], \;\;\; i=1,\ldots ,N, 
\label{eq1}
\end{eqnarray}
where ${\bf F}({\bf x})$ governs the dynamics of each individual oscillator, ${\bf H}({\bf x})$ is
a linear vectorial function, $\sigma$ is the overall coupling strength and the
$G=G_{ij}$ is the coupling matrix.
The rows of matrix $G$ 
have zero sum and this ensures that the completely synchronized state 
$\{ {\bf x}_i(t)={\bf s}(t), \forall i \; | \; \dot{\bf s}={\bf F}({\bf s}) \}$
is a solution of Eq.~(\ref{eq1}).
By means of the so called Master Stability Function approach, it is possible to
study the conditions under which such a state is stable \cite{SPRL}, i.e. 
the propensity for synchronization ($PFS$) of a given network.
\\         
In Ref.\cite{SPRL} it was shown that an enhancement 
in the $PFS$ can be achieved by exploiting
the information contained in the overall topology of the network.
This can be done through an opportune choice of the coupling matrix $G$ that makes
use of the \textit{load} concept and by scaling the coupling strenght $\sigma$ in Eq.~(\ref{eq1})
to the load of each link.
The load $l_{ij}$ of the link connecting nodes $i$ and $j$ is quantified by
the so called {\it edge betweenness}, i.e. the fraction of shortest paths
that are making use of that link. By means of this weighting procedure, that
clearly reflects the network structure at a global scale, Eq.~(\ref{eq1})
reads: 
\begin{eqnarray}
\dot{\bf x}_i&=&{\bf F}({\bf x}_i)-\frac{\sigma}{\sum_{j\in K_i} ~ l_{ij}^\alpha} 
\sum_{j\in K_i} l_{ij}^\alpha ~{\bf H}[{\bf x}_i - {\bf x}_j]~~\;\;\; i=1,\ldots ,N,  
\label{eq2}
\end{eqnarray}
where $\alpha$ is a real tunable parameter, and $K_i$ is the set of 
neighbors of node $i^{th}$.
For a given dynamical system ${\bf F}({\bf x}_i)$, for a given value of $\sigma$ and for a given network 
topology it is possible to find, by means of the
Master Stability Function approach \cite{SPRL}, what is the value of $\alpha_{best}$ providing
the best $PFS$ of the system. In practice, it is more convenient to put $\alpha=0$ and to find a value of
the coupling parameter $\sigma$ which would ensure a fully synchronized state for the oscillators network. 
\\
In Ref.\cite{BILPR} we used as dynamical system the so-called Opinion Changing
Rate (OCR) model \cite{ocr}, with an Heigselmann-Krause dynamics \cite{heg}, and we showed that, 
starting from a perfectly synchronized state for $\alpha=0$, if $\alpha$ is let to decrease 
from  $0$ to $-\infty$, the links with the higher load will be weighted
less and less with respect to the other links, thus inducing a progressive
desynchronization of the system in clusters of frequencies (dynamical clustering)
corresponding to different modules, or communities, of a given network.
Here we apply our analysis to real or trial networks using a system of 
chaotic R\"ossler oscillators and we study again the dynamical clustering
process as a function of decreasing values of $\alpha$, 
identifying a likely community subdivision of the networks
by looking to local or global maxima of the \textit{modularity} $Q$ \cite{BILPR}. 
The latter simply quantifies the 
degree of correlation between the probability of having an edge joining
two sites and the fact that the sites belong to the same community \cite{NG}, 
thus in general it makes sense to look for large values of $Q$. In fact we get
$Q=0$ if we consider the whole network as a single community or if we consider
a completely random network. On the other hand, for networks with an appreciable 
subdivision in classes, $Q$ usually falls in the range between $0.2$ and $0.7$. 
\\
The dynamics of a system of $N$ identical (three-dimensional) chaotic R\"ossler oscillators, 
defined over the nodes of a given network, is ruled by Eq.(\ref{eq2}), with
${\bf x}_i = (x_i, y_i, z_i)$, 
${\bf F}({\bf x}_i) = [-\omega y_i - z_i, \omega x_i + 0.165 y_i, 0.2 + z_i(x_i - 10)]$
and ${\bf H}({\bf x})= [x, 0, 0] $ (thus the coupling acts only on the $x$ variable). 
In other words we have the following equations of motion:
\begin{displaymath}
\dot{x}_i=-\omega y_i - z_i-\frac{\sigma}{\sum_{j\in K_i} ~ l_{ij}^\alpha} 
\sum_{j\in K_i} l_{ij}^\alpha ~(x_i - x_j)  
\end{displaymath}
\begin{equation}
\dot{y}_i=\omega x_i + 0.165 y_i
\end{equation}
\begin{displaymath}
\dot{z}_i=0.2 + z_i(x_i - 10) ~~~\;\;\;\;\;\; i=1,\ldots ,N,  
\label{rossler}
\end{displaymath}
Here $\omega$ is a common natural frequency associated
at each oscillator that, without loss of generality, we put equal to $1.0$. 
The load matrix $l_{ij}$ (the matrix of the edge betweennesses) is calculated once 
forever for the chosen network with a computational 
cost of $O(KN)$, $K$ being the total number of links. 
\begin{figure}
\includegraphics[height=.3\textheight]{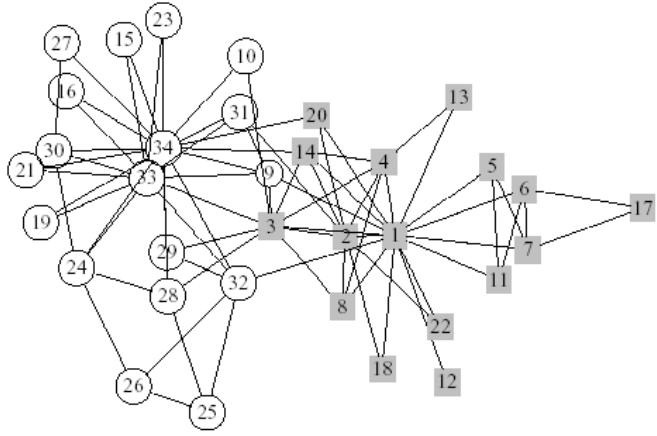}
\caption{The Karate Club network , with the two Zachary's communities 
identified by circles and squares. 
}
\end{figure}
\\
In order to evaluate the degree of synchronization of the R\"ossler system (\ref{rossler}) one 
has to calculate the order parameter 
$\Psi = \langle \frac{1}{N} \vert \sum_{i=1}^N e^{j\Phi_i (t)} \vert  \rangle_t$, 
where $\Phi_i (t)= arctan[\frac{y_i(t)}{x_i(t)}]$ indicates the istantaneous phase of
the $i$-th oscillator and $\langle . . .\rangle_t$ stays for a time average.
If all the oscillators rotate independently, no clusters exist and we have $\Psi \sim \frac{1}{\sqrt{N}}$.
On the contrary, if their motions are synchronized in phase, only one cluster exists and we obtain $\Psi \sim 1$.
Once a network is fixed, the first task is to find the value of the coupling parameter $\sigma$ 
providing a fully synchronized starting state for the R\"ossler oscillators at $\alpha=0$ (i.e. at $t=0$). 
Then, one can let $\alpha$ to decrease in time and study the dynamical clustering process
acting on the istantaneous phases $\Phi_i (t)$'s of the oscillators. Notice that these phases play here
the same role played by the istantaneous frequencies in the OCR model \cite{BILPR}:
in this case we call "cluster" a group of contiguous phases in the $\Phi$'s interval
(usually $[-3,3]$) separated by a distance of more than $0.02$ units. 
For each value of $\alpha$ a different configuration of clusters (corresponding to a given network structure) 
will appear and one has to calculate the corresponding modularity and select
the configuration with the best modularity score.
\\
In the following we will show in detail this process by putting the R\"ossler 
system over different real and trial networks.

\begin{figure}
\includegraphics[height=.25\textheight]{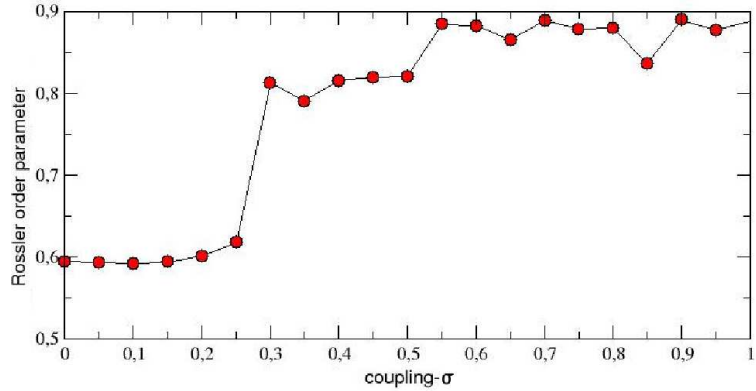}
\caption{Asymptotic R\"ossler order parameter 
(at $\alpha=0$) versus the coupling $\sigma$ for the Karate Club network. See text for further details.
}
\end{figure}

\subsection{Zachary's karate club}

As first example, in order to test our algorithm for finding community structures, we consider 
a real network, the well-known Karate Club network
analyzed by Zachary \cite{zachary}. It consists of $N=34$ individuals (nodes),-
 whose mutual friendship relations (expressed by $K=78$ edges)
have been carefully investigated over a period of two years. 
Due to contrasts between a teacher and the administrator of the club, 
the club splitted into two smaller communities. 
The corresponding network is presented in Fig.1, where squares and circles 
label the members of the two groups.
The 'circles' community has $18$ elements
(corresponding to nodes $9,10,15,16,19,21,23,24,25,26,27,28,29,30,31,32,33,34$)
while the 'squares' community has $16$ elements.
(nodes $5,6,7,11,17,1,2,3,4,8,$ $12,13,14,18,20,22$).
The $a$-$priori$ modularity of such a configuration results to be equal to $Q_{Z}\sim0.37$.
\\
Firstly, in Fig.2 we plot the behavior of the asymptotic R\"ossler order parameter 
(averaged over 10 events) versus the coupling strenght $\sigma$ for the Karate Club network with $\alpha=0$.
It results that, above $\sigma\sim0.6$, the system lies in the fully synchronized phase, 
thus in the following we reasonably set $\sigma=1.3$. Using such a value in the 
equations of motion (\ref{rossler}), we can now integrate them numerically and
study the dynamical clustering desynchronization process acting on the oscillators' phases.
The task of our algorithm is to extract the best modular structure of the network
by using only the information expressed by the edge betweenness of its links, which are calculated 
only once for the given network.
\\
The system starts in a perfectly synchronized state
for $\alpha_{start}=0$, i.e. $x_i(0)=y_i(0)=z_i(0)=0$ $\forall i$, 
thus $\Phi_i(0)=0$ $\forall i$ and only one cluster exists (choosing
a different starting value for $x_i,y_i,z_i$ does not change significantly the
conclusions shown in the following); 
then we decrease $\alpha$ {\it during a single run} as a function of time with a rate 
of $2$ time steps and a given constant decrement $\delta\alpha=0.0008$. 
Simultaneously, for each value of $\alpha(t)$ and for each configuration of
clusters, the modularity $Q$ is calculated.
%
\begin{figure}
\includegraphics[height=.35\textheight]{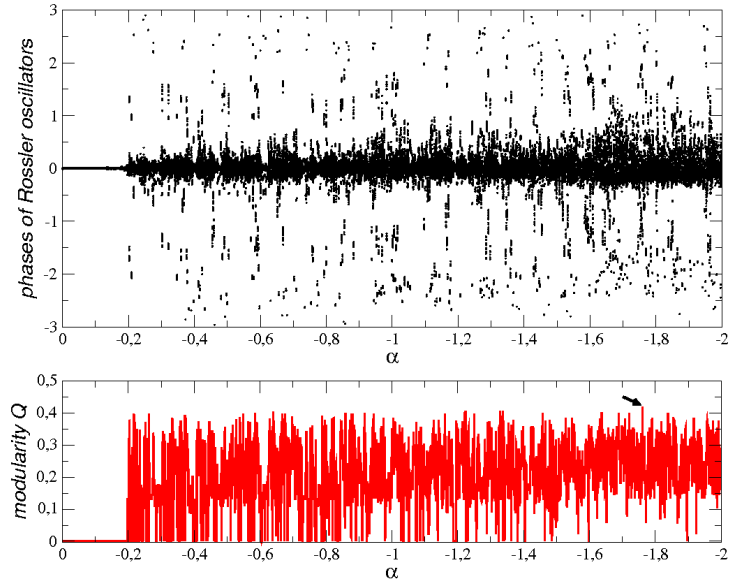}
\caption{Typical run for the Karate Club Network: time-evolution of both the R\"ossler's phases 
configurations (top panel) and the corresponding modularity (bottom panel) as a function of $\alpha(t)$.
See text for further details.
}
\end{figure}
%
\begin{table}
\begin{tabular}{lr}
\hline
cluster 1 (11 nodes) & 1,2,3,4,8,12,13,14,18,20,22\\
cluster 2 (12 nodes) & 9,10,15,16,19,21,23,27,29,30,31,32,33,34\\
cluster 3 (4 nodes) & 24,25,26,28\\
cluster 4 (2 nodes) & 27,30\\
cluster 5 (5 nodes) & 5,6,7,11,17\\
\hline
\end{tabular}
\caption{Clusters configuration with the best modularity score $Q_{best}=0.42$ for the Karate Club Network
(at $\alpha_{best}\sim-1.76$). See text for further details.}
\label{tab:1}
\end{table}
\\
In the top panel of Fig.3 we show the phases time evolution as a function of $\alpha(t)$
in a typical run with $\alpha$ going from $0$ to $-3$. 
In this figure $34$ points are plotted for each value of $\alpha(t)$, corresponding
to the istantaneous phases of the oscillators (from which the average istantaneous phase of
the system has been subtracted in order to have a symmetric plot). 
Correspondingly, in the bottom panel of Fig.3, the modularity $Q$ is also plotted 
as a function of $\alpha(t)$. 
Even if the system strongly oscillates during the desynchronization process 
(at variance with the much more stable behavior of the OCR system in \cite{BILPR}), 
clusters configurations (i.e. community structures of the 
underlying network) with very large values of modularity appear.
\\
In Table 1 we report the detailed clusters configuration corresponding 
to the maximum value of modularity, $Q_{best}=0.42$, found for $\alpha_{best}\sim-1.76$
(see the arrow in the bottom panel).
One can see that five clusters have been found: 
the first and the last ones, respectively made of $11$ and $5$ nodes (oscillators), 
if considered together, correspond to the first community of $16$ nodes 
(squares in the left panel of Fig.1) observed by Zachary,
while the sum of the remaining three clusters corresponds to tightly connected subgroups 
of the second community of $18$ nodes (circles in the left panel of Fig.1).
On the other hand, for several values of $\alpha$ in Fig.3 (e.g. for $\alpha=-0.538$ or $\alpha=-1.075$) 
the clusters configuration
corresponding to the two Zachary communities has been also recovered, but - as previously seen -, 
being its modularity smaller than $0.42$ (in fact $Q_{Z}\sim0.37$), the algorithm 
favours the five clusters configuration shown in Table 1 
(in other words, the Zachary community subdivision corresponds only to a local maximum 
of modularity for this particular network and not to a global one). 
\\
In conclusion, at least in so far as it concerns the Zachary Club network, the dynamical clustering algorithm 
based on the R\"ossler system seems to work very well, even if compared with the analogous results 
presented in \cite{BILPR}.
In fact it results to be at the same time very \textit{fast} (it extracts the best 
clusters configuration of the network versus $\alpha$ in a single run) 
and very \textit{sensitive} (it is able to recover  
community structures of the Zachary network with a great modularity).  
In the next section we will apply the same algorithm to another real network,
the Chesapeake Bay food web.

\subsection{Chesapeake Bay Food Web}

Another classical benchmark for the community identification algorithms 
is a food web of marine organisms living in the Chesapeake Bay, situated on the Atlantic
coast of the United States. 
This ecosystem was originally studied by
Baird and Ulanowicz \cite{baird}, who carefully investigated the predatory interactions 
between the most important taxa (species or groups of species)
and constructed a network of 33 vertex and 71 links.
We will consider here (as usually done in many papers \cite{BILPR,newgirv1,Eff}) 
its non-directed and non-valued version and
calculate first (only once before each simulation using the same network) 
the edge betweenness of each link. 
%
\begin{figure}
\includegraphics[height=.35\textheight]{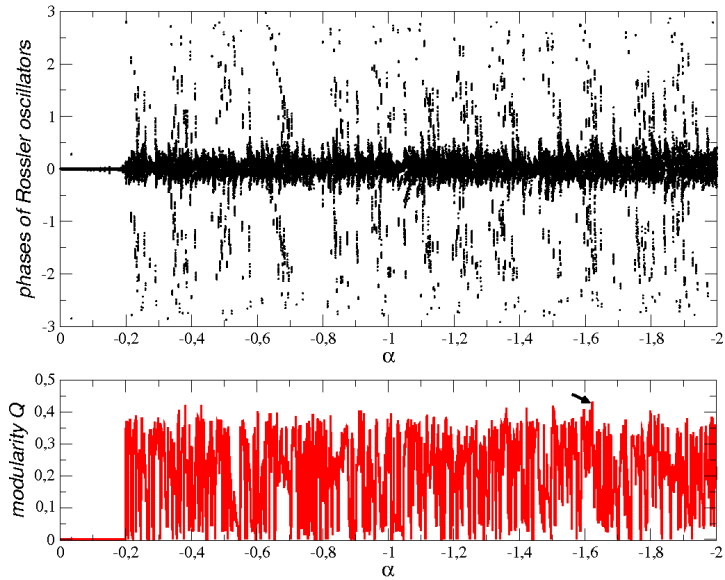}
\caption{Typical run for the Chesapeake Bay food web: time-evolution of both the R\"ossler's phases 
configurations (top panel) and the corresponding modularity (bottom panel) as a function of $\alpha(t)$.
}
\end{figure}
\\
In Fig.4 we show the result of a typical event obtained in a simulation performed 
with the same procedure described in the previous section and for a value of
the interaction strenght $\sigma=1$ (such that the R\"ossler's system would lie
in its synchronized phase for $\alpha=0$).
The system starts in a fully synchronized state (we set again $x_i(0)=y_i(0)=z_i(0)=0$ $\forall i$)
at $\alpha_{start}=0$ and evolves through decreasing values of $\alpha(t)$ 
(with a decrement $\delta\alpha=0.0008$), up to the value
$\alpha_{end}=-2$. The clusters evolution and the corresponding modularity $Q(t)$ are plotted as a function of $\alpha$ 
and, again, look very oscillating in time. The detailed configuration with the highest modularity peak (see the arrow
in the bottom panel) is reported in Table 2.
It consists of 6 clusters with a $Q_{max}=0.43$, obtained  for $\alpha_{AVT}=-1.62$,
and it is quite consistent with the corresponding results of \cite{BILPR,newgirv1,Eff}), 
where a main separation in two large communities had been found, according
to the distinction between pelagic organisms, which live near the surface or at middle depths (clusters n.1,2 and 4), 
and benthic organisms, which live near the bottom (clusters n.3,5 and 6).
\\
Such a result corroborates the good performance of the R\"ossler algorithm 
in the identification of community structures in real networks.
The next step will be to test this method on "ad hoc" trial networks
with a well known fixed community structure \cite{BILPR,NG}, 
in order to explore in deeper detail its sensitivity.

%
\begin{table}
\begin{tabular}{lr}
\hline
cluster 1 (10 nodes) & 3,14,15,16,18,25,26,27,28,29\\
cluster 2 (3 nodes) & 4,17,19\\
cluster 3 (1 nodes) & 30\\
cluster 4 (3 nodes) & 22,31,32\\
cluster 5 (14 nodes) & 1,2,7,8,9,10,11,12,13,20,21,23,24,33\\
cluster 6 (2 nodes) & 5,6\\
\hline
\end{tabular}
\caption{Clusters configuration with the best modularity score $Q_{best}=0.43$ for the Chesapeake Bay food web
(at $\alpha_{best}\sim-1.62$)}
\label{tab:2}
\end{table}

\subsection{Sensitivity test for ad hoc trial networks}

Typical trial networks are generated with $N = 128$ nodes and split into four
communities containing 32 nodes each. Pairs of nodes belonging to the
same community are linked with probability $p_{in}$, whereas pairs
belonging to different communities are joined with probability
$p_{out}$. The value of $p_{out}$ is taken so that the average number
of links a node has to members of any other community, i.e. $z_{out}$, can
be controlled. While $p_{out}$ (and therefore $z_{out}$) is varied
freely, the value of $p_{in}$ is chosen to keep the total average node
degree $k$ constant, and set to 16. As $z_{out}$ is increased
from $2$ (very well defined structures) to $8$ (bad defined structures), 
the communities become more and more diffuse and harder to
detect. Since the ``real'' community
structure is well known in this case, it is possible to measure the
number of nodes correctly classified by our method of community
identification (see for example \cite{BILPR}).
\\
We apply our algorithm to $10$ different sets of trial networks.
For each network, as done in the previous sections, after having calculated the load matrix 
$l_{ij}$, we integrate numerically the equations of motion (\ref{rossler}). 
Every time we start from a perfectly synchronized state for $\alpha_{start}=0$
and we analyze the desynchronization process when $\alpha(t)$ decreases in time 
(with a given constant decrement $\delta\alpha$) during a single run.
\begin{figure}
\includegraphics[height=.3\textheight]{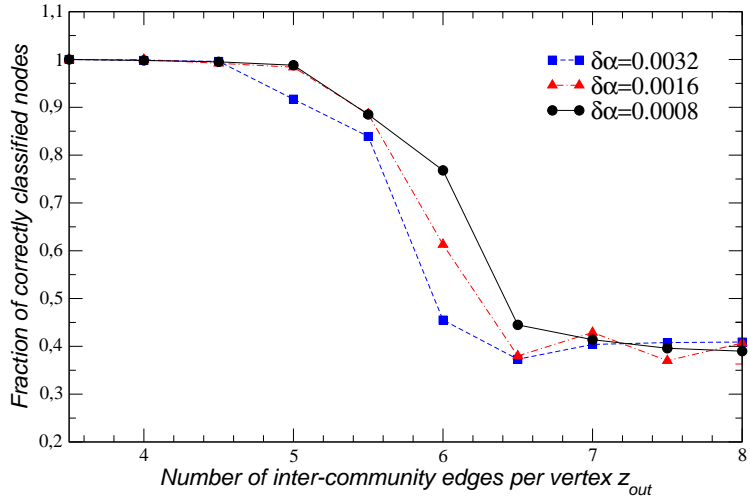}
\caption{Sensitivity curves of the dynamical clustering algorithm with R\"ossler system: the fraction of correctly classified nodes in a set of N=128 trial networks is reported as a function of $z_{out}$ for several increasing values of the decrement $\delta\alpha$.}
\end{figure}
In Fig.5 we plot the number of correctly classified nodes as a function of $z_{out}$,
averaged over the set of $10$ networks and for three increasing values of $\delta\alpha$.
The error bars (standard deviations) for each point is also reported.
As one can see, the smaller is $\delta\alpha$, the better is the result. 
However, in any case the sensitivity abruptly falls above $z_{out}=5$, staying around 
the $40\%$ of correctly identified nodes up to $z_{out}=8$.
Such a performance of the R\"ossler system is surely worse than that of 
the $OCR-HK$ system shown in \cite{BILPR}, but in any case  
it confirms the possibility of extending the dynamical clustering algorithm 
presented in \cite{BILPR} to other dynamical systems (also three-dimensional,
like in this case) with quite good results.

\section{Conclusion}

Summarizing, even if the global sensitivity of the dynamical clustering (DC) algorithm for 
the a system of chaotic R\"ossler oscillators seems to be not competitive with respect 
to those of other methods (see Refs.\cite{BILPR,comparing,gudkov}), on the other hand the results
here presented confirm that
(i) the DC algorithm is robust with respect to the change of the adopted dynamical system; 
(ii) the system of chaotic R\"ossler oscillators works quite well if applied 
to the real networks considered;
(iii) the DC algorithm is quite fast and needs only few runs of integration 
for each network (after the calculation of the corresponding load matrix $l_{ij}$)
\cite{BILPR,boc02,SPRL,NG}.



\end{document}